\RequirePackage{fix-cm}
\documentclass{svjour3}  
\usepackage[a4paper, total={6in, 8in}]{geometry} 
\smartqed  
\usepackage{graphicx}
\usepackage{dcolumn}
\usepackage{bm}
\usepackage{amsmath}

\usepackage{braket}
\usepackage[square,sort,comma,numbers]{natbib}

\def\bcen{\begin{center}}
\def\ecen{\end{center}}
\usepackage{epstopdf}
\journalname{Quantum Information Processing}
\begin{document}

\title{Enhancing phase sensitivity with number state filtered coherent states
}


\author{Nilakantha Meher$^a$         \and
        S. Sivakumar$^b$ 
}


\institute{$^a$Department of Physics, Indian Institute of Technology Kanpur, Kanpur, UP 208016, India.
\\
 $^b$ Division of Natural Sciences, Krea University, Andhra Pradesh 517646, India,\\ 
              \email{$^a$nilakantha.meher6@gmail.com}           
           \and
\email{$^b$sivakumar.srinivasan@krea.edu.in}
}

\date{Received: date / Accepted: date}

\maketitle

\begin{abstract}
Number state filtered coherent states are a class of nonclassical states  obtained by removing one or more number states from a coherent state. Phase sensitivity of an interferometer is enhanced if these nonclassical states are used as input states.  The optimal phase sensitivity, which is related to the quantum Cramer-Rao bound (QCRB) for the input state, improves beyond the standard quantum limit.   It is argued that removal of more than one suitable number state leads to better phase sensitivity.  As an important limiting case in this  context, the  even and odd coherent states, where the odd and even number states are filtered from coherent state respectively, are considered.  The optimal phase sensitivity for these limiting cases  equals  that of the squeezed vacuum.    It is observed that the improvement in  phase sensitivity is not in direct proportion to the nonclassicality of the input states.
\end{abstract}

\section{Introduction}
Interferometery is the primary tool for estimating change of phase of optical fields  \cite{Hariharan,Giovannetti,Barnett, Abramovici}.  It relies on the changes in the interference pattern when a phase object is introduced in one of the arms of an interferometer, for instance, the Mach-Zehnder interferometer (MZI). An important requirement in phase measurement is to achieve  precision as close to the theoretical bound given by the laws of quantum mechanics, which are much lower than the classical limit on interferometric phase measurements. Phase sensitivity of an interferometric configuration depends on the states of the light at the two input ports.\\

Among the pure states of light, the coherent states are important as they possess features such as factorizable coherence functions, localized phase space distribution, minimal quadrature fluctuations, etc \cite{Glau, Sudar, Glau2}.  Aptly, these states are classified as the most classical among the quantum states.     With the classical states, \textit{i.e.,} coherent state,  as one of the input states and vacuum as the other input,  phase sensitivity is limited by the standard quantum limit (SQL), \textit{i.e.,} $\Delta\theta_{\text{SQL}}=1/\sqrt{N}$, where $N$ is the average number of photons in the coherent state \cite{Gerry, Demkowicz}. In this arrangement, sensitivity can be enhanced by increasing the average number of photons. But higher intensity of the field increases the radiation pressure which decreases sensitivity \cite{Caves, Caves2}. However, SQL is not the lowest bound for phase  measurements.  If  suitable nonclassical states are  used as input states sensitivity can be improved \cite{Giovannetti2011, Escher2011}. For instance, squeezed vacuum (SV) and coherent state in the input of MZI beats the standard quantum limit \cite{Caves}. The optimal phase shift that can be resolved using these input states is $\Delta\theta=e^{-r}/\sqrt{N}$, where $r$ is the squeezing parameter \cite{Lang}. The minimum phase shift that can be detected in an interferometer is bounded by the Heisenberg limit (HL) \cite{Caves2, Gerry}, \textit{i.e.} $\Delta\theta_{\text{HL}}=1/N$, where $N$ is the average number of photons input to the interferometer.  Enhancement of phase sensitivity beyond SQL can be achieved using nonclassical states in the input.   For instance, NOON state \cite{Boto, Jonathan, Hayashi}, entangled coherent state \cite{Joo, Joo2, Liu}, pair coherent state \cite{GerryMimih}, twin Fock state \cite{Campos, Lang2}, squeezed thermal state and even or odd coherent states \cite{TanFranco}, coherent state and number state \cite{Birrittella}, two mode squeezed states \cite{Bondurant}, etc  improve  phase sensitivity of interferometer.   \\

Another interesting case is to replace the vacuum port with a coherent state input so that the input is a double coherent state, \textit{i.e.,} $\ket{\alpha}\ket{\beta}$.   The optimal phase sensitivity that can be achieved is $\Delta\theta=1/\sqrt{|\alpha|^2+|\beta|^2}$ \cite{Ataman}.  If $\alpha=\beta$, then $\Delta\theta=1/\sqrt{2N}$, $N$ being the average number of photons in any one of the inputs.  Substituting the vacuum with a coherent state in the input improves the phase sensitivity.  However, it does not go below SQL. This arises from the fact that the classical input states  remain  classical at the output of MZI.\\   


It is of interest to construct input states that overcome SQL.  In this article, we investigate the phase sensitivity of a special class of non-classical states called the number state filtered coherent states (NFCS) \cite{MeherNSFS}. Quantum Cramer-Rao bound (QCRB) gives the optimal phase sensitivity achievable for a given input state \cite{Cramer, Rao, Kok}. Hence, QCRB is used to quantify the achievable phase sensitivity. Using NFCS and coherent state, Heisenberg limit is asymptotically reached for large average photon number. Filtering a single number state from coherent state improves the phase sensitivity. It can be improved further by filtering more than one number state from the coherent state. We show that the odd coherent state (OCS) and even coherent state (ECS), where even and odd photon number states are filtered out, perform as well as SV for phase measurement.\\

This article is organized as follows: A brief review of calculation of quantum Fisher information (QFI) and QCRB is given in Sec. \ref{SecQFI}. Phase sensitivity of single number state filtered coherent state is discussed in Sec. \ref{SingleNSFS}. In Sec. \ref{multiNSFS}, the optimal phase sensitivity achievable using multi-number states filtered coherent state such as ECS and OCS is presented. We discuss the role of nonclassicality of the quantum state on phase sensitivity in Sec. \ref{Nonclassicality}. Finally, we summarize the results in Sec. \ref{Summary}.

\section{Quantum Fisher information}\label{SecQFI}
In this section, a brief overview on quantum Fisher information  of the input state to the MZI is presented. Interestingly, QFI is related to the quantum Cramer-Rao bound \cite{Helstrom, Kok,Jarzyna, Lang2}, which provides a lower bound in phase sensitivity.  
Recent studies have employed  QFI  in various contexts such as an hierarchical ordering  of quantum states \cite{Erol}, characterizing the non-Markovianity of open quantum processes \cite{Lu}, understanding the thermal entanglement \cite{Ozaydin2}, etc.   This measure has also been  investigated in studying the  robustness of  W-states \cite{Ozaydin} and  GHZ states \cite{Ma} under dissipation and decoherence.   For optimizing the phase sensitivity of an interferometer in the presence of dissipation or decoherence, Heisenberg-Langevin approach or quantum master equation approach can be implemented to calculate QFI \cite{Wang3}.  However, in this work we have assumed that dissipation and decoherence are negligible and are neglected. \\
 
\begin{figure}
\centering
\includegraphics[width=8cm,height=5.5cm]{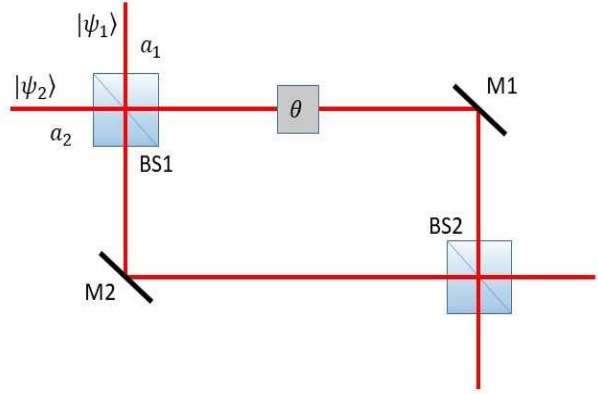}
\caption{Sketch of a Mach-Zhender interferometer. Two states $\ket{\psi_1}$ and $\ket{\psi_2}$ are the input to the ports 1 and 2 respectively. The phase element in the upper arm introduces a relative phase $\theta$ between the paths. }
\label{MZI}
\end{figure}

Let the input state to the interferometer be a product state of the form $\ket{\psi_{in}}=\ket{\psi_1}\ket{\psi_2}$, where $\ket{\psi_1}$ and $\ket{\psi_2}$ are the input states to the ports 1 and 2 respectively as indicated  in Fig. \ref{MZI}. Relevant to phase sensitivity of MZI is the QFI of the state that results after the action of the first beam splitter and phase shift due to the phase object. The unitary operator to represent the action of the first beam splitter (50:50) is $U_{BS}=e^{-i\frac{\pi}{4}(a_1^\dagger a_2+a_1 a_2^\dagger)}$ and for  the relative phase shift the operator is  $U(\theta)=e^{-i\theta(a_1^\dagger a_1-a_2^\dagger a_2)/2}$.\\

Defining 
\begin{align}
J_x=\frac{1}{2}(a_1^\dagger a_2+a_1 a_2^\dagger),\\
J_y=\frac{-i}{2}(a_1^\dagger a_2-a_1 a_2^\dagger),\\
J_z=\frac{1}{2}(a_1^\dagger a_1-a_2^\dagger a_2),
\end{align}
The unitary operators are recast as $U_{BS}=\exp \left(\frac{-iJ_x\pi}{2}\right)$ and $U_\theta=\exp(-i\theta J_z)$ respectively.\\

Quantum Fisher information for a pure state $\ket{\psi_{in}}$ is \cite{Braunstein,Kok,Jarzyna, Lang2}
\begin{align}\label{FisherInfPureState}
F_{Q}(\theta)=4\{\langle \psi'(\theta)|\psi'(\theta)\rangle-|\langle \psi'(\theta)|\psi(\theta)\rangle|^2\},
\end{align}
where
\begin{align}\label{state}
\ket{\psi(\theta)}&=U(\theta) U_{BS}\ket{\psi_{in}}=e^{-i\theta J_z}\exp \left(-\frac{iJ_x\pi}{2}\right)\ket{\psi_{in}},
\end{align} 
and
\begin{align}\label{statederivative}
\ket{\psi'(\theta)}=\frac{\partial}{\partial\theta}\ket{\psi(\theta)}=-ie^{-i\theta J_z}J_z\exp\left(-\frac{iJ_x\pi}{2}\right)\ket{\psi_{in}}.
\end{align}

Using the Eqns. \ref{state} and \ref{statederivative}, QFI given in Eqn. \ref{FisherInfPureState} becomes
\begin{align}\label{FisherInfJy}
F_{Q}(\theta)=4[\langle \psi_{in}|J_y^2|\psi_{in}\rangle-\langle \psi_{in}|J_y|\psi_{in}\rangle^2],
\end{align}
which is proportional to the variance of $J_y$ in the input state $\ket{\psi_{in}}$. Here we have used the identity 
\begin{align}
\exp\left(\frac{iJ_x\pi}{2}\right)J_z\exp\left(-\frac{iJ_x\pi}{2}\right)=J_y.
\end{align}
Using $J_y=\frac{-i}{2}(a_1^\dagger a_2-a_1 a_2^\dagger)$, Eqn. \ref{FisherInfJy} can be written as
\begin{align}\label{FisherAnnihilationCreation}
F_{Q}(\theta)&=\langle a_1^\dagger a_1\rangle+\langle a_2^\dagger a_2\rangle-\langle a_1^{\dagger 2}a_2^2\rangle-\langle a_1^{2}a_2^{\dagger 2}\rangle \nonumber\\+2 &\langle a_1^\dagger a_1 a_2^\dagger a_2\rangle+\langle a_1^\dagger a_2\rangle^2+ \langle a_1 a_2^\dagger\rangle^2-2\langle a_1^\dagger a_2\rangle \langle a_1 a_2^\dagger\rangle,
\end{align}
where $\langle~\rangle$ implies expectation value in the input state $\ket\psi_{in}$.\\

The phase shift that can be resolved by a given input state in an interferometric arrangement is bounded by \cite{Caves,Braunstein,Kok,Jarzyna, Lang2}
\begin{align}\label{QCRB}
\Delta\theta\geq \frac{1}{\sqrt{F_{Q}(\theta)}},
\end{align}
the quantum Cramer-Rao bound \cite{Cramer,Rao}. The optimal phase shift $\Delta\theta_{QCRB}=1/\sqrt{F_{Q}(\theta)}$. This relationship implies that for better phase sensitivity, the states input to the MZI should have larger $F_{Q}(\theta)$.\\

We assume that the input to the port 2 is a coherent state $\ket{\beta}$ throughout the discussion. For this choice,
\begin{align}\label{QFICoh}
F_{Q}(\theta)&=\langle a_1^\dagger a_1\rangle+|\beta|^2+2|\beta|^2(\langle a_1^\dagger a_1\rangle-\langle a_1^\dagger \rangle \langle a_1\rangle)-|\beta|^2(\langle a_1^{\dagger 2}\rangle+\langle a_1^{ 2}\rangle-\langle a_1^{\dagger }\rangle^2-\langle a_1\rangle^2).
\end{align}
The expectation values of operators for the input port 1 appearing in Eqn. \ref{QFICoh} are to be calculated in $\ket{\psi_1}$.
\section{Single number state filtered coherent state} \label{SingleNSFS}
In this section, we study the phase sensitivity of single number state filtered coherent state (SNFCS) in terms of quantum Cramer-Rao bound. SNFCS is defined as \cite{MeherNSFS}
\begin{align}\label{SNSFS}
\ket{\psi(\alpha,m)}=\frac{e^{-|\alpha|^2/2}}{N_m}\sum_{n=0,n\neq m}^{\infty}\frac{\alpha^n}{\sqrt{n!}}\ket{n},
\end{align}  
where $N_m=\sqrt{1-e^{-|\alpha|^2}\frac{|\alpha|^{2m}}{m!}}$. This definition implies that the number state $\ket{m}$ is absent from the coherent state. Hence, the probability of detecting $m$ photons in $\ket{\psi(\alpha,m)}$ is zero.\\
\subsection{Phase sensitivity with SNFCS and coherent state input}
In order to investigate the advantages of using SNFCS, consider the input state $\ket{\psi_{in}}=\ket{\psi(\alpha,m)}\ket{\beta}$, wherein the vacuum state in the input port has been replaced with $\ket{\psi(\alpha,m)}$. From Eqn. \ref{QFICoh}, QFI associated with this input is
\begin{align}\label{QFINSFS1}
F_{\text{S}}(\theta)&=\langle a_1^\dagger a_1\rangle+|\beta|^2+2|\beta|^2(\langle a_1^\dagger a_1\rangle-\langle a_1^\dagger \rangle \langle a_1\rangle)-|\beta|^2(\langle a_1^{\dagger 2}\rangle+\langle a_1^{ 2}\rangle-\langle a_1^{\dagger }\rangle^2-\langle a_1\rangle^2),
\end{align}
where
\begin{align}
\langle a_1^\dagger a_1 \rangle &=\frac{1}{N_m^2}[|\alpha|^2-m(1-N_m^2)],\\
\langle a_1\rangle &=\alpha+\frac{m}{\alpha^*}-\frac{m}{\alpha^* N_m^2},\\
\langle a_1^2\rangle &=\alpha^2+\frac{m(m-1)}{\alpha^{*2}}-\frac{m(m-1)}{\alpha^{*2} N_m^2}.
\end{align}
The expectation values of operators relevant for the input port 1 are calculated in $\ket{\psi(\alpha,m)}$.\\

QCRB for the aforementioned input state is
\begin{align}\label{QCRBsNSFS}
\Delta\theta_{\text{S}}=\frac{1}{\sqrt{F_{\text{S}}(\theta)}},
\end{align}
and the corresponding HL and SQL are
\begin{align}
\Delta\theta_{\text{HL}}=\frac{1}{\frac{1}{N_m^2}[|\alpha|^2-m(1-N_m^2)]+|\beta|^2},
\end{align}
and
\begin{align}
\Delta\theta_{\text{SQL}}=\frac{1}{\sqrt{\frac{1}{N_m^2}[|\alpha|^2-m(1-N_m^2)]+|\beta|^2}}.
\end{align}

\begin{figure}
\centering
\includegraphics[width=9cm,height=8cm]{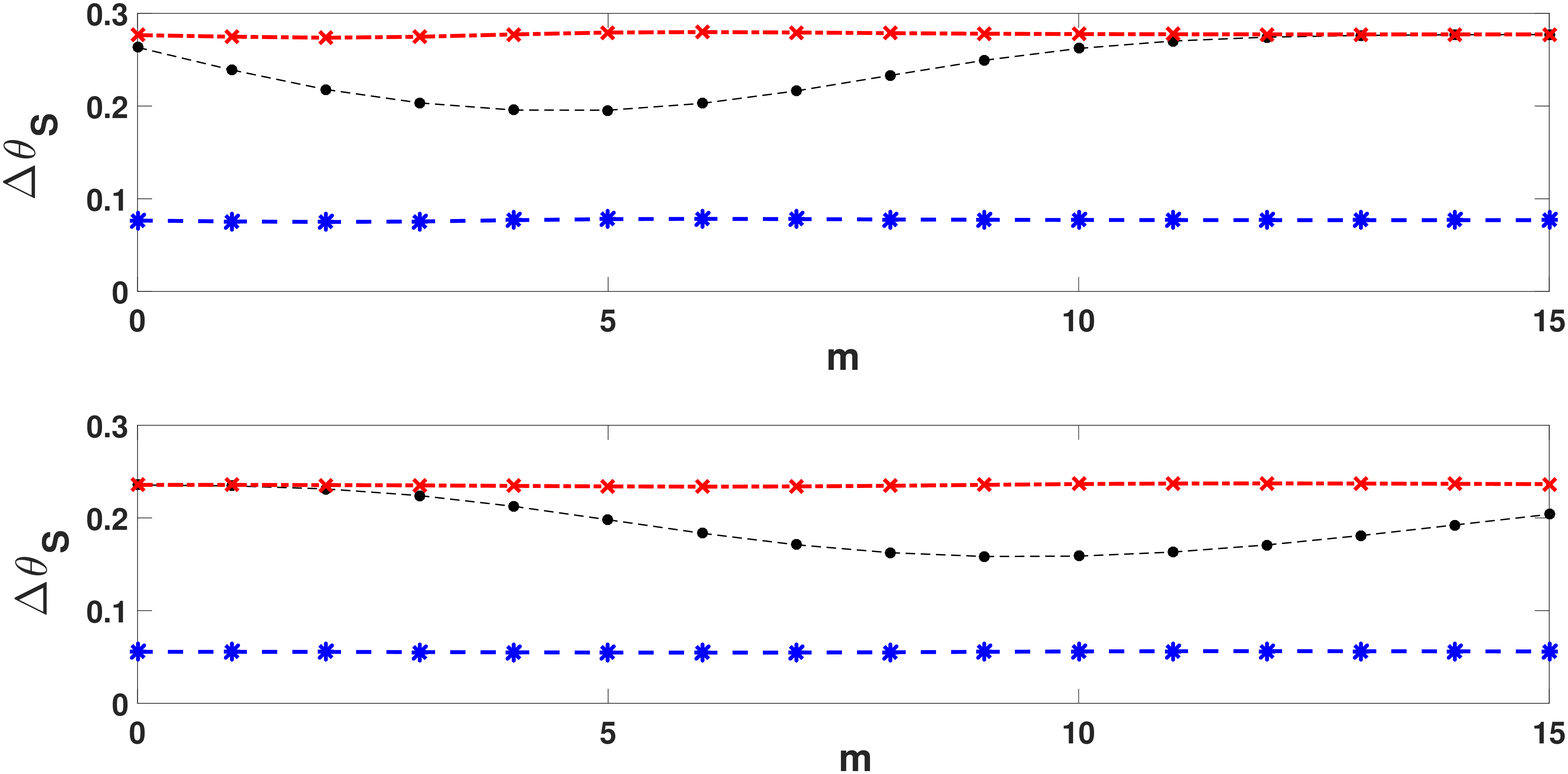}
\caption{Quantum Cramer-Rao bound $\Delta\theta_{\text{S}}$ (larger dot) for the input state $\ket{\psi_{in}}=\ket{\psi(\alpha,m)}\ket{\beta}$ as a function of $m$. Here $(a)\alpha=2,\beta=3$ and $(b)\alpha=3,\beta=3$.  It is compared with $\Delta\theta_{\text{HL}}$ (star) and $\Delta\theta_{\text{SQL}}$ (cross). Phase sensitivity beats standard quantum limit.}
\label{DthetaNSFSm}
\end{figure}

In Fig. \ref{DthetaNSFSm},  $\Delta\theta_{\text{S}}$ is shown as a function of $m$.   For comparison, $\Delta\theta_{\text{HL}}$ and $\Delta\theta_{\text{SQL}}$ are also shown.
It is to be noted that the optimal phase sensitivity $\Delta\theta_{\text{S}}$ surpasses SQL if SNFCS is used instead of vacuum state as the input in MZI. This indicates that filtering a number state from coherent state enhances phase sensitivity in an interferometric phase measurement. Moreover, maximum sensitivity is achieved when $|\alpha|^2= m$, as seen in Fig. \ref{DthetaNSFSm}.  On the other hand, if $|\alpha|^2>>m$ or $|\alpha|^2<<m$, $\Delta\theta_{\text{S}}$ reaches $\Delta\theta_{\text{SQL}}$. In this limit $|\langle \psi(\alpha,m)|\alpha\rangle|^2 \approx 1$, which implies $\ket{\psi(\alpha,m)}\approx \ket{\alpha}$ \cite{MeherNSFS}. \\  

\begin{figure}
\centering
\includegraphics[width=9cm,height=5.5cm]{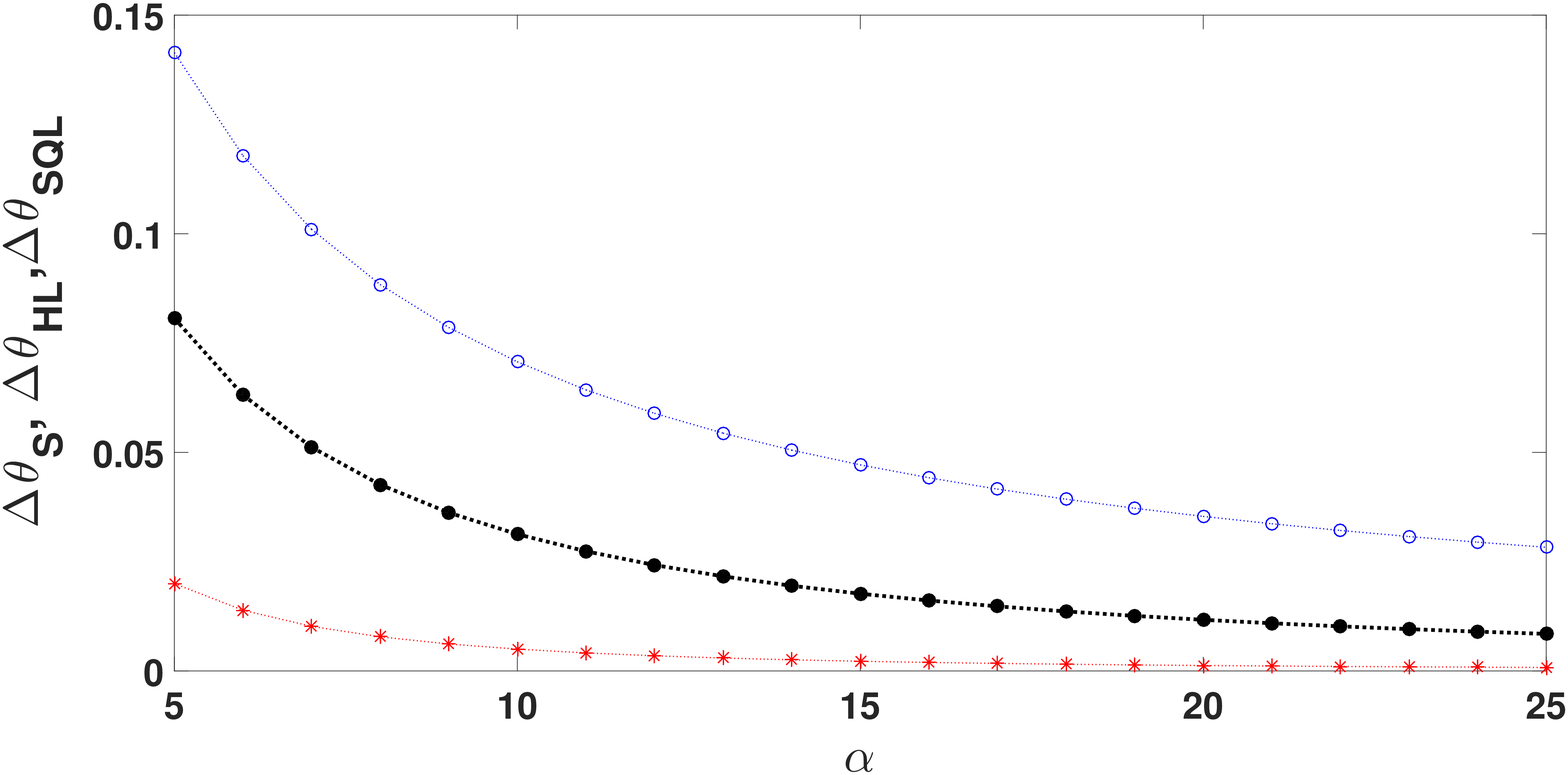}
\caption{Quantum Cramer-Rao bound $\Delta\theta_{\text{S}}$ for the input state $\ket{\psi_{in}}=\ket{\psi(\alpha,m)}\ket{\beta}$ (middle curve).  Result is compared with Heisenberg limit $\Delta\theta_{\text{HL}}$ (lower curve) and SQL $\Delta\theta_{\text{SQL}}$ (upper curve). Here we assume $\alpha=\beta$, and $m=|\alpha|^2$, $\alpha$ and $\beta$ are assumed as real.}
\label{HeisenbergLimitSNSFS}
\end{figure}

Interestingly, for $|\alpha|^2=|\beta|^2>>1$ and $m=|\alpha|^2$, $\Delta\theta_{\text{S}}$ given in Eqn. \ref{QCRBsNSFS} becomes 
\begin{align}
\Delta\theta_{\text{S}}=\frac{\sqrt{2}}{N}\sqrt{\frac{N_m^2}{1-N_m^2}},
\end{align}
which shows the Heisenberg scaling $\Delta\theta_{\text{S}}\propto 1/N$, as can be seen in Fig. \ref{HeisenbergLimitSNSFS}. Here $N=|\alpha|^2+|\beta|^2$,  which is the total average number of photons at the MZI input. It is to be noted that the average number of photons in $\ket{\psi(\alpha,m)}$ is $|\alpha|^2$ when $m=|\alpha|^2$.\\

\begin{figure}
\centering
\includegraphics[width=9cm,height=5.5cm]{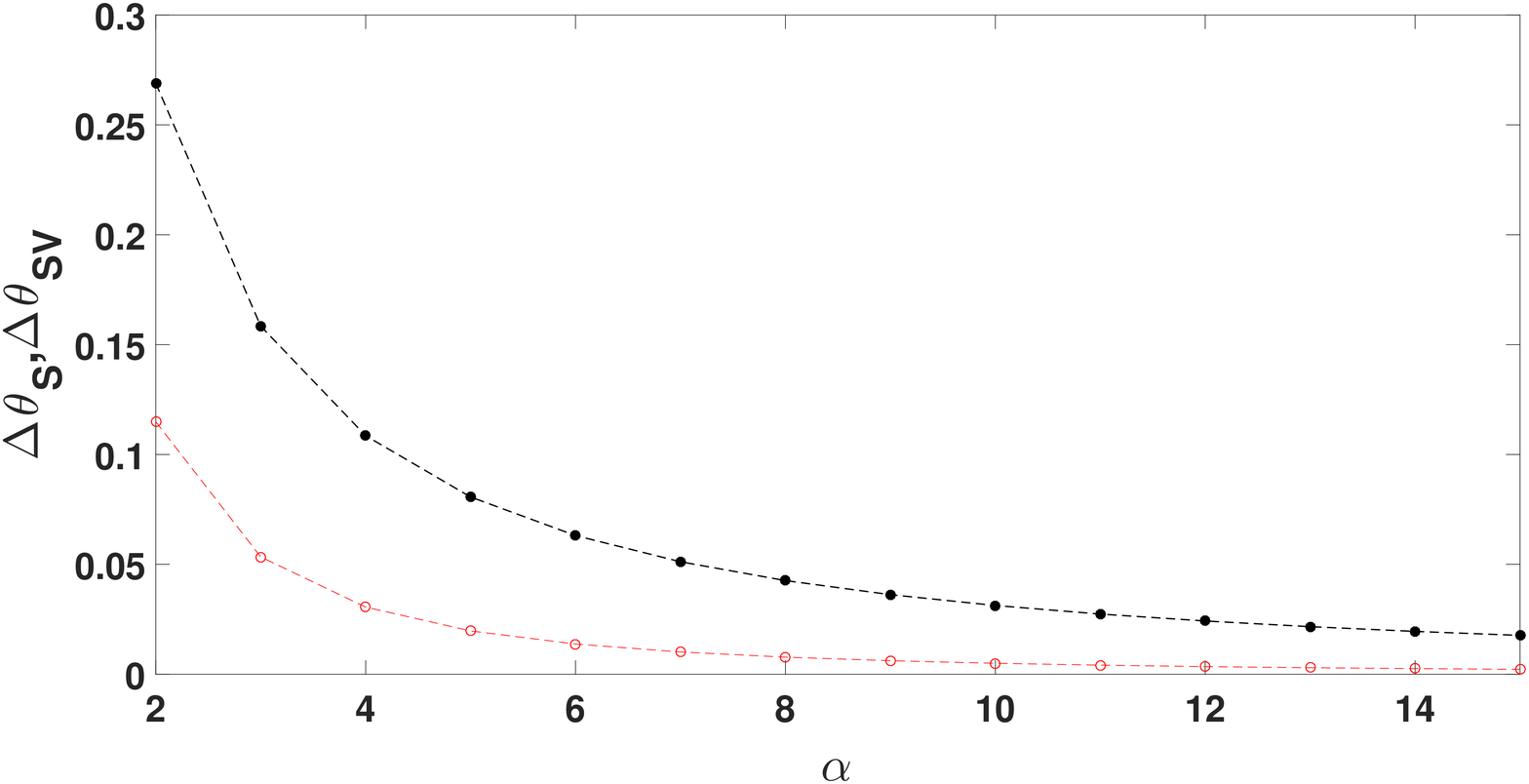}
\caption{ $\Delta\theta_{\text{S}}$ and $\Delta\theta_{\text{SV}}$ are shown as a function of $\alpha$ respectively for the input states $\ket{\psi(\alpha,m)}\ket{\beta}$ (upper curve) and $\ket{\xi}\ket{\beta}$ (lower curve).  Here we assume $\alpha=\beta$, $m=|\alpha|^2$ and $r=\sqrt{\sinh^{-1}|\alpha|^2}$. $\alpha$ and $\beta$ are assumed as real. }
\label{NSFSvsSQZplot}
\end{figure}

It is known that SV and coherent state at the input of a MZI achieves  optimal phase sensitivity. As a comparison, we show QCRB for $\ket{\psi(\alpha,m)}\ket{\beta}$ and $\ket{\xi}\ket{\beta}$ in Fig. \ref{NSFSvsSQZplot}, where $\ket{\xi}$ is SV. For the input state $\ket{\xi}\ket{\beta}$, QCRB is \cite{Caves,Caves2} 
\begin{align}\label{DthetaSQZ}
\Delta\theta_{\text{SV}}=\frac{1}{\sqrt{|\beta|^2e^{2r}+\sinh^2r}},
\end{align}
where $r$ is the squeezing parameter chosen in such a way that $|\alpha|^2=\sinh^2r$. This condition ensures that the average number of photons in $\ket{\xi}$ and $\ket{\psi(\alpha,m)}$ are equal if $m=|\alpha|^2$. As can be seen in Fig. \ref{NSFSvsSQZplot}, $\Delta\theta_\text{SV}<\Delta\theta_\text{S}$, which indicates that the performance of SV is better than that of SNFCS.
\section{Multi-number states filtered coherent state}\label{multiNSFS} 
If more than one number state are filtered from a coherent state, the resultant state is a multi-number states filtered coherent state (MNFCS). In the number state basis
\begin{align}\label{mNSFS}
\ket{\psi(\alpha,\{m\})}\propto e^{-|\alpha|^2/2}\sum_{n=0,n \not\in \{m\}}^{\infty}\frac{\alpha^n}{\sqrt{n!}}\ket{n},
\end{align}
where the set $\{m\}$ contains more than one integer. If $\{m\}$ has only a single member, this state reduces to the state given in Eqn. \ref{SNSFS} which is the SNFCS.\\ 

Two special classes of states, namely, even coherent state and odd coherent state belong to this category \cite{Dodonov, Buzek}. If the set $\{m\}$ contains only odd (even) integers, then the resultant state is even (odd) coherent state. Number state representation of ECS and OCS are
\begin{align}
\ket{\alpha,+}\propto e^{-|\alpha|^2/2}\sum_{n=0}^\infty \frac{\alpha^{2n}}{\sqrt{(2n)!}}\ket{2n},
\end{align}
and
\begin{align}
\ket{\alpha,-}\propto e^{-|\alpha|^2/2}\sum_{n=0}^\infty \frac{\alpha^{(2n+1)}}{\sqrt{(2n+1)!}}\ket{2n+1},
\end{align}
respectively. These states are nonclassical \cite{Dodonov,Buzek}.
\subsection{Phase sensitivity using MNFCS and coherent state}
Let the input state to MZI is $\ket{\psi(\alpha,\{m\})}\ket{\beta}$. QCRB for this input can be calculated using the Eqn. \ref{QFICoh} where the expectation values are to be calculated in the state $\ket{\psi(\alpha,\{m\})}$. We denote the QCRB for this input as $\Delta\theta_{\text{M}}$.\\  

Define
\begin{align}
\Theta=\frac{\Delta\theta_{\text{M}}}{\Delta\theta_{\text{HL}}}, 
\end{align}
to quantify the effectiveness of the input state. The case $\Theta=1$ corresponds to the ultimate phase sensitivity achievable. Here $\Delta\theta_{\text{HL}}=1/[\langle a_1^\dagger a_1\rangle +|\beta|^2]$, where $\langle a_1^\dagger a_1\rangle=\bra{\psi(\alpha,\{m\})}a_1^\dagger a_1 \ket{\psi(\alpha,\{m\})}$ is the average number of photon in MNFCS .\\

\begin{figure}[h!]
\centering
\includegraphics[width=9.5cm,height=5.8cm]{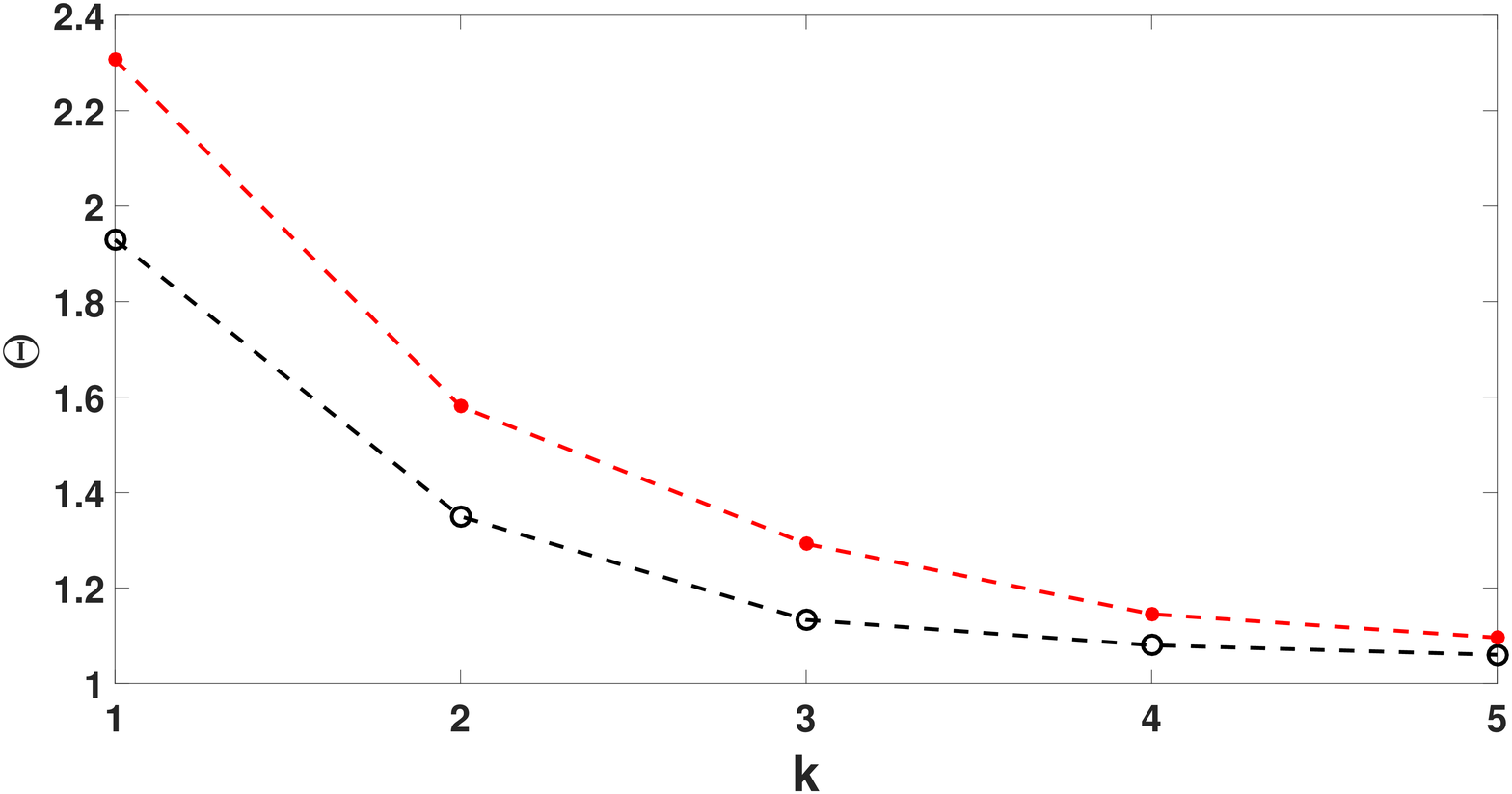}
\caption{ $\Theta$ is shown as a function of number of elements $(k)$ in the set $\{m\}$ for the state $\ket{\psi(\alpha,\{m\})}\ket{\beta}$ where $\alpha=3e^{i\pi/2}$ (circle) and $4e^{i\pi/2}$ (dot). We set $\beta=|\alpha|$.  }
\label{ThetaVsk}
\end{figure}

The ratio $\Theta$ depends on the choice of $\{m\}$. The optimal choice is the set which yields the lowest $\Delta\theta_{M}$ for a given $\alpha$ and $\beta$. The ratio $\Theta$ for  the input $\ket{\psi(\alpha,\{m\})}\ket{\beta}$ is shown in Fig. \ref{ThetaVsk} as a function of $k$. Here $k$ denotes the number of Fock state filtered, equivalently, number of  elements in the set $\{m\}$.  For example, SNFCS corresponds to $k=1$ as the set $\{m\}$ contains only one integer.  It is to be noted that as $k$ increases, \textit{i.e.,} number of Fock state filtered from the coherent state increases, $\Theta$ decreases. This signifies an improvement of phase sensitivity. Hence, filtering number states from coherent state enhances phase sensitivity. \\

QCRB for $\ket{\alpha,+}\ket{\beta}$ as the input state is
\begin{align}\label{DthetaECS}
\Delta\theta_{+}=\frac{1}{\sqrt{\langle a_1^\dagger a_1\rangle+|\beta|^2(1+2\langle a_1^\dagger a_1\rangle-2|\alpha|^2\cos2\theta)}},
\end{align}
where $\theta=\text{arg}(\alpha)$ and
\begin{align}
\langle a_1^\dagger a_1\rangle=\bra{\alpha,+}a_1^\dagger a_1 \ket{\alpha,+}=|\alpha|^2\tanh |\alpha|^2,
\end{align}
is the average number of photons in ECS.\\

A similar expression for QCRB for the input state $\ket{\alpha,-}\ket{\beta}$, \textit{i.e.,} $\Delta\theta_{-}$  is given by substituting $\bra{\alpha,+}a_1^\dagger a_1 \ket{\alpha,+}$ by $\bra{\alpha,-}a_1^\dagger a_1 \ket{\alpha,-}$ in Eqn. \ref{DthetaECS}. Here
\begin{align}
\bra{\alpha,-}a_1^\dagger a_1 \ket{\alpha,-}=|\alpha|^2\coth|\alpha|^2,
\end{align}
is the average number of photons in OCS. For $|\alpha|>>1$, $\bra{\alpha,+}a_1^\dagger a_1 \ket{\alpha,+} \approx  \bra{\alpha,-}a_1^\dagger a_1 \ket{\alpha,-} \approx |\alpha|^2$.\\

\begin{figure}[h!]
\centering
\includegraphics[width=9cm,height=5.5cm]{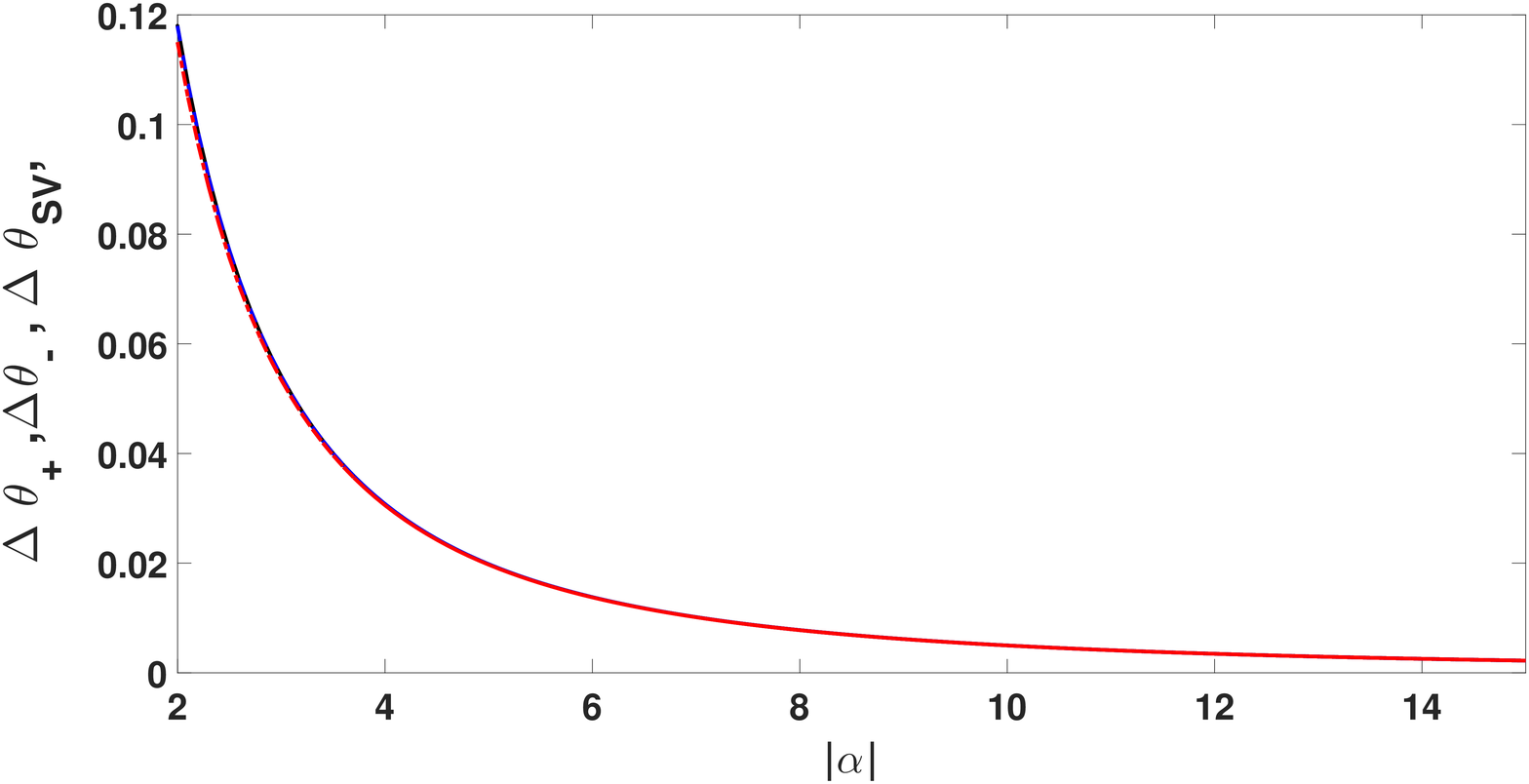}
\caption{Quantum Cramer-Rao bounds $\Delta\theta_{+}$ (continuous), $\Delta\theta_{-}$ (dashed) and $\Delta\theta_{\text{SV}}$ (dot-dashed) are shown as a function of $|\alpha|$.  Here we assume $\beta=|\alpha|$, $\theta=\pi/2$. We choose $\alpha$ and $r$ such that $\bra{\alpha,+}a_1^\dagger a_1 \ket{\alpha,+} \approx  \bra{\alpha,-}a_1^\dagger a_1 \ket{\alpha,-} \approx \bra{\xi}a_1^\dagger a_1 \ket{\xi}$. }
\label{EVSOCSSQZQCRB}
\end{figure}

Fig. \ref{EVSOCSSQZQCRB} shows $\Delta\theta_{SV}, \Delta\theta_{-}$ and $\Delta\theta_{+}$ as a function of $|\alpha|$ for $\theta=\pi/2$. We choose $\alpha$ and $r$ such that $\bra{\alpha,+}a_1^\dagger a_1 \ket{\alpha,+} \approx  \bra{\alpha,-}a_1^\dagger a_1 \ket{\alpha,-} \approx \bra{\xi}a_1^\dagger a_1 \ket{\xi}$. All the three curves coincide for large average photon number. For $|\alpha|^2>>1$,
\begin{align}
\Delta\theta_{+}\approx\Delta\theta_{-}\approx \frac{1}{\sqrt{|\alpha|^2+|\beta|^2+4|\alpha|^2|\beta|^2}}\approx\frac{1}{\sqrt{|\beta|^2e^{2r}+\sinh^2r}},
\end{align}
which are equal to the $\Delta\theta_{\text{SV}}$ given in Eqn. \ref{DthetaSQZ}. Hence,  ECS and OCS perform as good as SV for phase measurement. 
\section{Nonclassicality and phase sensitivity}\label{Nonclassicality}
Phase sensitivity of an interferometric setup using classical inputs can not surpass standard quantum limit. It requires one of the inputs to be non-classical to go beyond SQL. A pertinent question that arises is whether a maximally nonclassical state shows maximum phase sensitivity. \\ 
\begin{figure}[h!]
\centering
\includegraphics[width=9cm,height=6cm]{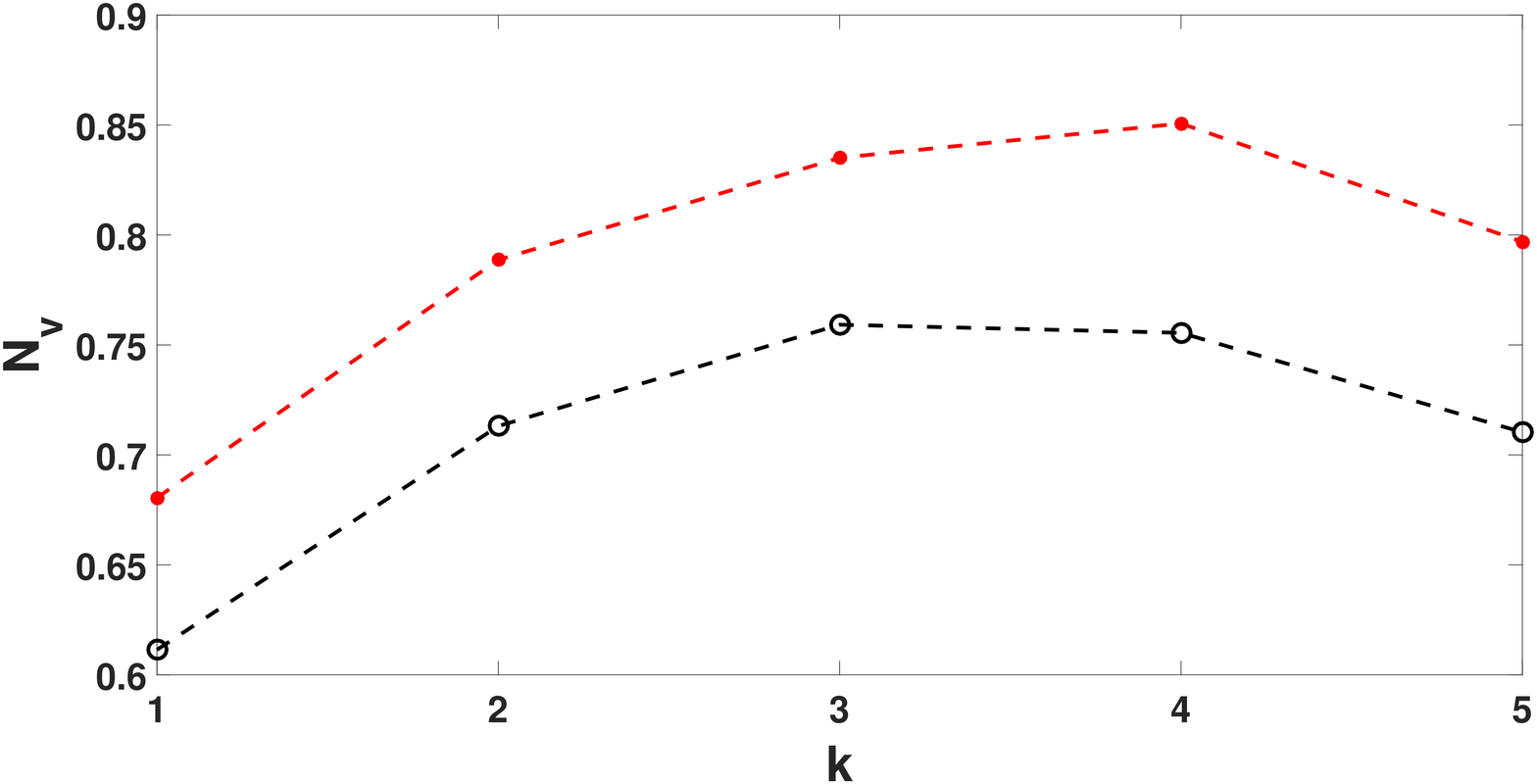}
\caption{Negativity of Wigner function for the states $\ket{\psi(\alpha,\{m\})}$ which achieve minimum $\Delta\theta_{\text{M}}$. Here $\alpha=3 e^{i\pi/2}$(circle) and $4e^{i\pi/2}$ (dot).  }
\label{NegativityVsk}
\end{figure}

 In this section, we discuss the role of nonclassicality on phase measurement using number state filtered coherent states. Nonclassicality of a state $\ket{\phi}$ is quantified in terms of the volume $(N_v)$ of the negative portion of Wigner function  $W_{\ket{\phi}}(\chi)$ \cite{Kenfack},
\begin{align}
N_v=\int\int\frac{|W_{\ket{\phi}}(\chi)|-W_{\ket{\phi}}(\chi)}{2}d\chi'd\chi'', 
\end{align} 
where $\chi'$ and $\chi''$ are the real and imaginary part of $\chi$. For a given $\ket{\phi}$,
\begin{align}
W_{\ket{\phi}}(\chi)=\frac{2}{\pi}\sum_{n=0}^{\infty}(-1)^n|\bra{\phi}D(\chi)\ket{n}|^2,
\end{align}
is the Wigner function for the state $\ket{\phi}$ \cite{Wigner,Gerry}.
Here $D(\chi)=\exp(\chi a^\dagger-\chi^* a)$ is the displacement operator \cite{Glau,Gerry}.\\

Negativity of Wigner function for those states which achieve minimum $\Delta\theta_{\text{M}}$ is shown as a function of $k$ in Fig. \ref{NegativityVsk}.  Phase sensitivity increases as $k$ increases (refer Fig. \ref{ThetaVsk}) whereas the nonclassicality initially increases and then decreases as shown in Fig. \ref{NegativityVsk}. Hence, there is no guarantee that increasing nonclassicality will enhance phase sensitivity.
\section{Summary}\label{Summary}
Coherent state and the vacuum as input states limit the phase sensitivity of Mach-Zehnder interferometer to the standard quantum limit  which is inversely proportional to the square root of the mean number of photons in the input state.   Surpassing this limit requires replacement of the vacuum by a nonclassical state.  Phase sensitivity, quantified in terms of quantum Cramer-Rao bound, shows improvement beyond the standard quantum limit if number state filtered coherent states are used instead of the vacuum.    Phase sensitivity increases further if many number states are removed from the coherent state.   As an extension, the even and odd coherent states also classified as number state filtered coherent states.  These states, with  infinite number of states filtered, show significant improvement in phase sensitivity.  For equal average number of photons,  even and odd coherent states’ performance is as good as that of the  squeezed vacuum as far as phase measurements are concerned. Though phase sensitivity increases with removing more number states from the coherent state, it is seen that nonclassicality does not increase monotonically. We will extend this work to study the effects of dissipation and decoherence on phase sensitivity to account for non-ideal mirrors and beam-splitters.



\section{Acknowledgement}
NM acknowledges Indian Institute of Technology Kanpur for postdoctoral fellowship.



%
%

\end{document}